\documentclass[11 pt, a4paper]{article}
\usepackage{jheppub}
\usepackage{amsmath}

\newcommand{\Tr}{{\rm Tr\;}}

\usepackage{graphicx}
\usepackage{amsmath}
\usepackage{amsfonts}
\usepackage{epsfig}
\usepackage{slashed}
\usepackage{braket}

\usepackage{comment}
\usepackage{cases}

\def\nn{\nonumber}
\def\bec{\begin{center}}
\def\eec{\end{center}}
\def\beq{\begin{equation}}
\def\eeq{\end{equation}}
\def\bea{\begin{eqnarray}}
\def\eea{\end{eqnarray}}

\def\ahat{\hat{a}}
\def\bhat{\hat{b}}
\def\chat{\hat{c}}
\def\dhat{\hat{d}}
\def\ehat{\hat{e}}

\newcommand{\cF}{{\cal F}}
\newcommand{\cFb}{{\overline{\cal F}}}
\newcommand{\cD}{{\cal D}}
\newcommand{\cDb}{{\overline{\cal D}}}
\newcommand{\cQ}{{\cal Q}}
\newcommand{\cU}{{\cal U}}
\newcommand{\cN}{{\cal N}}
\newcommand{\cUb}{{\overline{\cal U}}} 

\title{Lattice ${\cal N}=4$ super Yang-Mills at Strong Coupling}
\author[a]{Simon Catterall,}
\author[b]{Joel Giedt,}
\author[a] {Goksu Can Toga}

\affiliation[a]{Department of Physics, Syracuse University, Syracuse, NY 13244, USA }
\affiliation[b]{Department of Physics and Astronomy, RPI, Troy, NY 12180, USA}

\abstract{In this paper we 
present results from numerical simulations of ${\cal N}=4$ super Yang-Mills for two
color gauge theory over a wide range  of 't Hooft coupling $0<\lambda\le 30$ using
a supersymmetric lattice action \cite{Catterall:2009it}.
Numerical study of this lattice theory
has been stymied until recently by both sign problems and the occurrence
of lattice artifact phases at strong coupling. We have recently developed a new action that appears
capable of solving both problems. The resulting
action possesses just $SU(2)$ rather than $U(2)$ gauge symmetry. 
By explicit computations of the fermion Pfaffian we present evidence that the theory 
possesses no sign problem and exists in a single phase out to arbitrarily strong coupling. Furthermore, preliminary work shows that the logarithm of
the supersymmetric Wilson loop varies as the square root of the 't Hooft coupling $\lambda$
for large $\lambda$
in agreement with holographic predictions.}
\preprint{preprint}

\begin{document}
\maketitle

\section{Introduction}
In this paper we use numerical simulation to explore the phase structure and Wilson loops of a lattice
formulation of ${\cal N}=4$ super Yang-Mills. The lattice action is a generalization of the formulation
described in \cite{Catterall:2009it}.
The theory preserves both $SU(N)$ gauge invariance, a $S^4$ point group symmetry associated with the underlying  $A_4^*$ lattice
and most importantly a single exact supersymmetry. 

The original supersymmetric lattice formulation of ${\cal N}=4$ SYM 
has been the subject of a great deal of both numerical and analytical work  \cite{Catterall:2011pd,Catterall:2012yq,Catterall:2013roa,Catterall:2014vka}.
General arguments have been put forward that the theory should approach the continuum
${\cal N}=4$ theory after tuning a single marginal operator. However, after
some initial successes the numerical
work has been handicapped by two problems: the existence of a chirally broken
phase  for 't Hooft couplings $\lambda > 4$ and the observation of a sign problem
which develops in a similar region of coupling \cite{Schaich:2018mmv}. While these problems are not
present in dimensionally reduced versions of the theory \cite{Anagnostopoulos:2007fw,Hanada:2008gy,Catterall:2008yz,Catterall:2009xn,Catterall:2010fx,Hanada:2016zxj,Berkowitz:2016jlq,Catterall:2017lub,Rinaldi:2017mjl}, they have prevented the
systematic investigation of the four dimensional theory. The chirally broken phase has
been linked to the condensation of monopoles associated with the $U(1)$ sector of
the theory \cite{Catterall:2014vga}.

In this paper we show that the situation is markedly improved if one
adds a new operator to the lattice action which preserves the $S^4$ symmetry and
exact supersymmetry
but explicitly breaks the $U(N)$ gauge symmetry down to $SU(N)$. 

\section{Review of the old supersymmetric construction}
We start from the supersymmetric lattice action appearing in \cite{Catterall:2009it}.
\beq
S=\frac{N}{4\lambda} \cQ \sum_{x}\Tr \left(\chi_{ab}\cF_{ab}+\eta \cDb_a\cU_a+\frac{1}{2}\eta d\right)+S_{\rm closed}\eeq
where the lattice field strength
\beq\cF_{ab}(x)=\cU_a(x)\cU_b(x+\ahat)-\cU_b(x)\cU_a(x+\bhat)\eeq where $\cU_a(x)$
denotes the complexified gauge field living on the lattice link running from $x\to x+\ahat$ where $\ahat$ denotes one
of the five basis vectors of the underlying $A_4^*$ lattice.
Similarly
\beq\cDb_a \cU_a=\cU_a(x)\cUb_a(x)-\cUb_a(x-\ahat)\cU_a(x-\ahat).\eeq
The five fermion fields $\psi_a$, being superpartners of the (complex) gauge fields, live on the 
corresponding links, while
the ten fermion fields $\chi_{ab}(x)$ are associated with new face links running
from $x+\ahat+\bhat\to x$. The scalar fermion $\eta(x)$ lives on the lattice site $x$ and is
associated with the conserved supercharge $\cQ$
which acts on the fields in the following way
\begin{align}
\cQ\, \cU_a&\to \psi_a\nn\\
\cQ\, \psi_a&\to0\nn\\
\cQ\, \eta&\to d\nn\\
\cQ\, d&\to 0\nn\\
\cQ\, \chi_{ab}&\to \cFb_{ab}\nn\\
\cQ\, \cUb_a&\to 0
\end{align}
Notice that $\cQ^2=0$ which guarantees the supersymmetric invariance of the
first part of the lattice action. The auxiliary site field $d(x)$ is needed for nilpotency of $\cQ$ offshell.
The second term $S_{\rm closed}$ is given by
\beq
S_{\rm closed}=-\frac{N}{16\lambda}\sum_x \Tr \epsilon_{abcde}\chi_{ab}\cDb_c\chi_{de}\eeq
where the covariant difference operator acting on the fermion field $\chi_{de}$ takes the form
\beq
\cDb_c\chi_{de}(x)=\cUb_c(x-\chat)\chi_{de}(x+\ahat+\bhat)-\chi_{de}(x-\dhat-\ehat)\cUb_c(x+\ahat+\bhat)\eeq

The latter term can be shown to be supersymmetric via an exact lattice
Bianchi identity $\epsilon_{abcde}\cDb_c \chi_{de}=0$. 
Carrying out the $\cQ$ variation and integrating out the auxiliary field $d$ we obtain the
supersymmetric lattice action $S=S_b+S_f$ where
\beq
S_b=\frac{N}{4\lambda} \sum_x\Tr\left( \cF_{ab}\cFb_{ab}+\frac{1}{2}\Tr (\cDb_a \cU_a)^2\right)\eeq
and
\beq
S_f=-\frac{N}{4\lambda}\sum_x \left(\Tr\chi_{ab}\cD_{\left[a\right.}\psi_{\left. b\right]}+
\Tr\eta \cDb_a\psi_a\right)\eeq
In the continuum this action can be obtained by discretization of
the Marcus or GL twist of $\cN=4$ Yang-Mills 
but in flat space is completely equivalent to it. In the continuum the twist is done as a prelude
to the construction of a topological quantum field theory but in the context of
lattice supersymmetry it is merely used as a change of variables that
allows for discretization while preserving a single exact supersymmetry. The twisting removes the spinors
from the theory replacing them by the antisymmetric tensor fields $\eta,\psi_a,\chi_{ab}$ which appears as components
of a K\"{a}hler-Dirac field. The latter is equivalent at zero coupling to a (reduced) staggered field
and hence describes four physical Majorana fermions in the continuum limit - as required for $\cN=4$ Yang-Mills. The
twisting procedure also
packs the six scalar fields of the continuum theory together with the four gauge fields
into five complex gauge fields corresponding to the lattice fields $\cU_a$.

As described above, the discrete
theory is defined on a somewhat exotic lattice - $A_4^*$. This admits a larger
set of rotational symmetries than a hypercubic lattice and this fact plays 
a role in controlling the renormalization
of the theory. Finally, to retain exact supersymmetry all fields reside in
the algebra of the gauge group -- taking their values in the adjoint representation of $U(N)$:
$f(x)=\sum_{A=1}^{N^2} T^A f^A(x)$ with $\Tr (T^A T^B)=-\delta^{AB}$.

Ordinarily this would be incompatible with
lattice gauge invariance because the measure would not be gauge invariant for
link based fields. However, in this $\cN=4$ construction the problem is evaded since the fields are 
complexified which ensures that the Jacobians that arise after gauge transformation of $\cU$ and $\cUb$ cancel.\footnote{Actually one should qualify this statement. While
the complexified bosonic measure is invariant under lattice $U(N)$
gauge transformations it is more subtle to show that the fermion measure is
invariant when the fermions reside on links. We shall show that this
issue is completely evaded in the theory with $SU(N)$ gauge invariance}

However this restriction to the algebra does pose a further problem. Ordinarily
the naive continuum limit is obtained by expanding the group elements about the
identity $U_a(x)=I+aA_a(x)+\ldots$. The presence of the unit matrix in this
expansion is what gives rise to hopping terms in the lattice theory
and derivative operators in the continuum limit. If the gauge fields live in the group the unit matrix
arises naturally on expanding the exponential but with fields valued in
the algebra it is less clear how such an expansion arises. The saving grace is
to notice that the gauge fields take their values in $GL(N,C)$  
so that this term can arise by giving a vacuum
expectation value to the {\it imaginary} part of the trace mode of the field.
Typically this is accomplished by adding to the supersymmetric action
a new  term of the form
\beq
S_{\rm mass}=\mu^2\sum_x \Tr\left(\cUb_a(x)\cU_a(x)-I\right)^2\eeq
While this breaks the exact supersymmetry softly all counter terms induced by this
breaking will have couplings that are multiplicative in $\mu^2$  and hence vanishing
as $\mu^2\to 0$. Notice also that this term also generates masses for the scalar
fields in the theory and hence also regulates the usual flat directions of SYM theory.

\section{The new action}
It has been observed that for couplings $\lambda>2$ the action described in
the previous section undergoes a phase transition to a regime in
which both the Polyakov line and the Wilson loop fall abruptly toward zero.
Associated with this is a growth in the density of lattice $U(1)$ monopoles \cite{Catterall:2014vka}.
These features are inconsistent with the expected superconformal phase of
$\cN=4$ Yang-Mills. Actually,
in pure compact QED in four dimensions, this monopole transition is a well known
lattice artifact. Various efforts have been made over the intervening years to
remove this monopole phase - typically this has been done by adding supersymmetric
or non-supersymmetric terms to the action that force the determinant of the plaquette
operator to unity. Such a procedure retains the full $U(N)$ gauge symmetry but restricts the
fluctuations of the field strength in the $U(1)$ directions.
The supersymmetric plaquette term introduced in \cite{Catterall:2015ira}
represents the best of these approaches but can only allow simulation up to
$\lambda\sim 6.0$. It also suffers from a sign problem for $\lambda>4$ \cite{Schaich:2018mmv} - that is, the
Pfaffian arising after fermion integration, exhibits strong phase fluctuations which prohibit
Monte Carlo sampling.

Here we explore an approach in which a new supersymmetric
term is introduced which drives the determinant of each individual gauge link to unity. 
The new
term takes the form
\beq
\frac{N}{4\lambda}\kappa\cQ \sum_{x,a} \Tr(\eta)\left({\rm Re\,det\,} (U_a(x)-1\right))\eeq
After $\cQ$ variation and integration over $d$ this modifies the second term
in the bosonic action $S_b$ to:
\beq
\frac{N}{4\lambda}\sum_{x,a} \frac{1}{2}\Tr\left(\cDb_a \cU_a(x)+\kappa{\rm Re\,det} (U_a(x))I_n\right)^2\eeq
where $I_N$ denotes the $N\times N$ unit matrix.
A corresponding new fermion term is generated
\beq
\delta S_f=-\frac{N}{8\lambda}\kappa\sum_{x,a}\Tr(\eta){\rm det\,} (U_a(x))\Tr(U_a^{-1}(x)\psi_a(x))\eeq
The new term has the effect of suppressing the $U(1)$ phase
fluctuations of the complex gauge links that were the origin of the monopole problem.
Of course this term explicitly breaks the $U(1)$ gauge symmetry. However since
the $U(1)$ is simply a decoupled free theory in the continuum limit this
should cause no real harm since  $SU(N)$ gauge invariance is preserved.
Indeed, close to the continuum limit, it should be apparent that the new terms merely
generate mass terms for the trace components of the fields.

In the original theory the gauge links were valued in $GL(N,C)$. After this term is added the moduli space of the theory is reduced to $SL(N,C)$. Notice
that since any matrix in $SL(N,C)$ can be written as the exponential
of a traceless matrix the presence of this
term  {\it guarantees} that gauge links can be expanded about the unit
matrix for vanishing values of the lattice spacing. In this light the remaining rationale for keeping $S_{\rm mass}$ is simply to lift the usual $SU(N)$ flat directions.
Indeed, as the reader will see, for most of our results $\mu^2$ is taken very small.

\begin{figure}[htbp]
\centering
\includegraphics[width=0.75\textwidth]{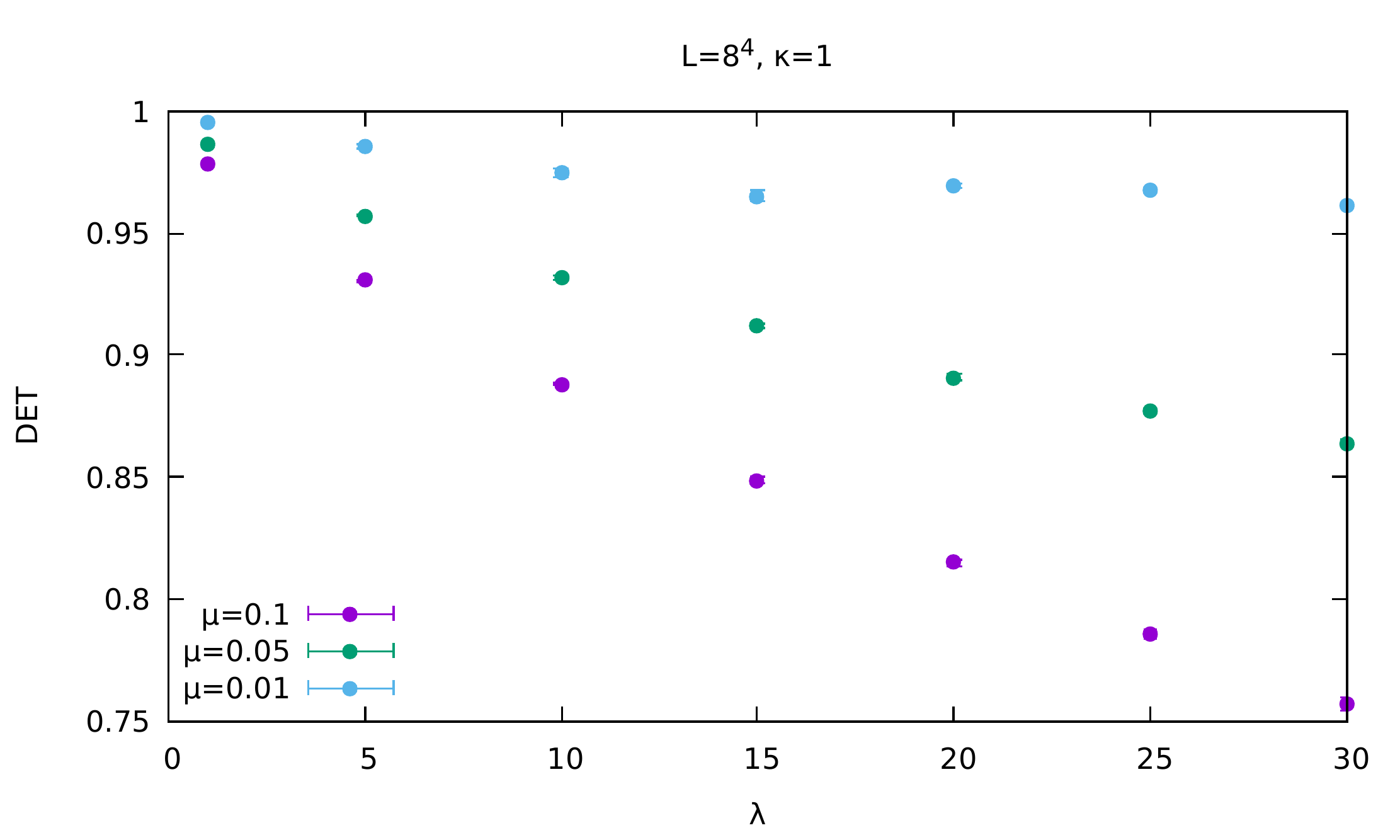} 
\caption{\label{det}Expectation value of the determinant vs $\lambda$  for $8^4$ lattices at $\mu=0.1,0.05,0.01$}
\end{figure}

The breaking of $U(1)$ gauge invariance also clarifies a delicate issue concerning the
invariance of the fermion measure in the original formulation. Consider the integration
measure for the five link fermions $\prod_{x,a} d\psi_a(x)$ in
the $U(N)$ theory. Under a gauge transformation
$\psi_a(x)\to G(x)\psi_a(x)G^\dagger(x+\ahat)$ this measure transforms by a non-trivial Jacobian corresponding to the
product of the determinants of the gauge factors $G(x)$ and $G^\dagger(x+\ahat)$. On the torus one can arrange an ordering of
the fermion fields in the path integral measure such that
these factors will cancel out along closed loops but this will not be possible for
all lattice topologies. Thus the question of the invariance of the measure under the full $U(N)$ group is a delicate one.
However these problems are completely avoided if $G$ is restricted to lie in
just $SU(N)$ as in the new action and the fermion measure is then unambiguously defined for an arbitrary lattice.

Of course the main question is whether such a term is effective at eliminating the
monopole phase seen at strong coupling. In the next section we shall show evidence
that this is true and at least in the case of 2 colors we 
see no sign of phase transitions out to arbitrarily large 't Hooft coupling.

\section{Phase structure}
\begin{figure}[htbp]
\centering
\includegraphics[width=0.75\textwidth]{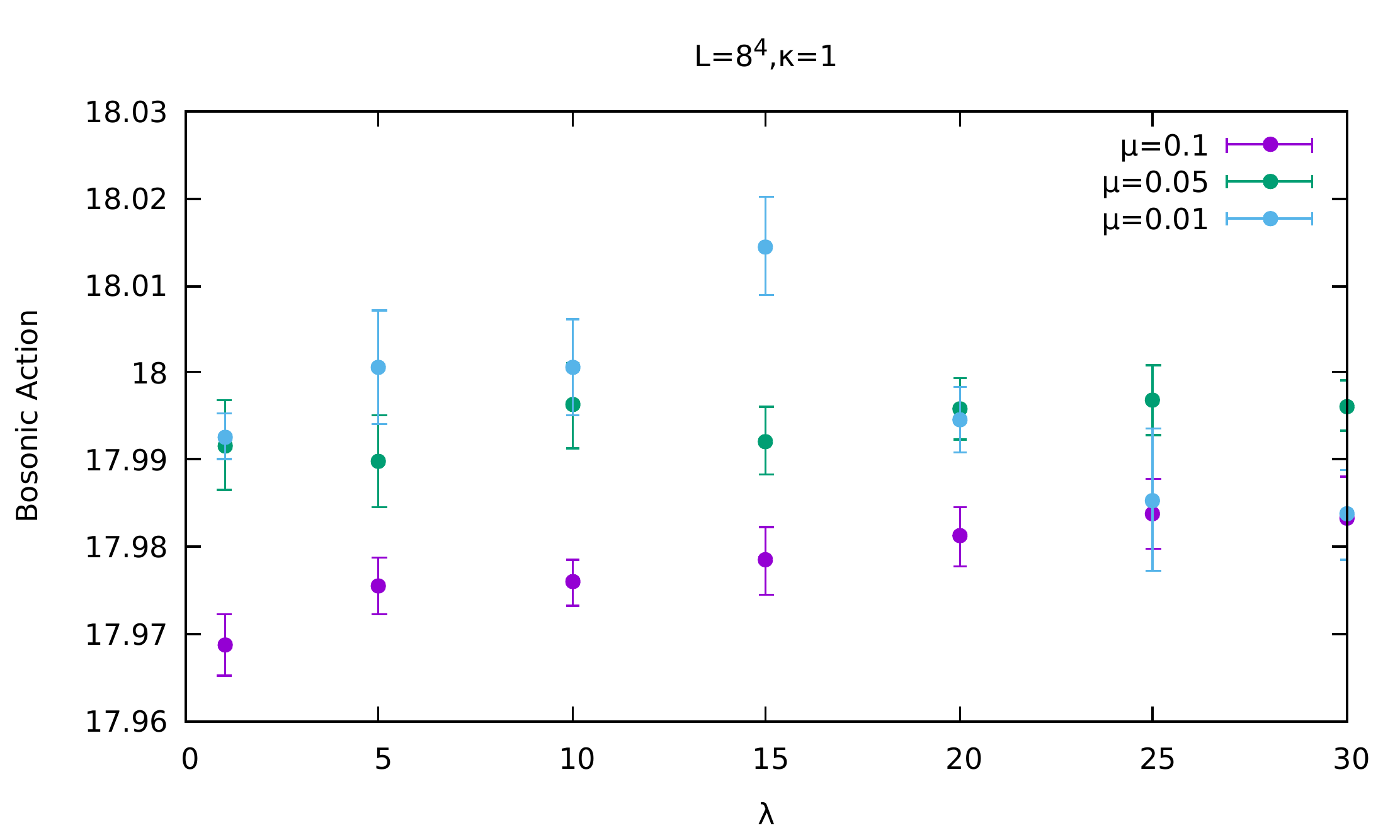} 
\caption{\label{baction}Expectation value of the bosonic action vs $\lambda$ for $8^4$ lattices at $\mu=0.1,0.05,0.01$}
\end{figure}

Our simulations utilize the rational hybrid Monte Carlo (HMC) algorithm where the
Pfaffian resulting from the fermion integration is replaced by
\beq{\rm Pf}\,(M)=\left({\rm det}    M^\dagger M\right)^{\frac{1}{4}}\eeq
where $M$ is the fermion
operator. Notice that this representation neglects any Pfaffian phase which is a key issue which we will return to later.
Typical ensembles used in our analysis consist of $5000$ HMC trajectories with $1000-2000$ discarded for
thermalization. Errors are assessed using a jackknife procedure using $20-40$ bins.

As a test of the new action we first plot the expectation value of the link determinant  as
a function of 't Hooft coupling. We show results in fig~\ref{det} for $8^4$ lattices at $\mu=0.1,0.05,0.01$. Clearly the expectation value is  close to
unity out to very large $\lambda$ provided $\mu^2$ is small enough confirming that we have effectively reduced the
gauge fields to $SU(2)$.  We note that we scan out to $\lambda=30$ in order to go beyond the self-dual point $\lambda_\text{SD}=4 \pi N = 8 \pi$.

In fig.~\ref{baction} we plot the expectation value of the bosonic action as a function of $\lambda$ for $8^4$ lattices at $\mu=0.1,0.05,0.01$. This expectation value
can be calculated exactly by exploiting the (almost) $\cQ$-exact nature of the lattice
action and yields $\frac{1}{V}<S_b>=\frac{9N^2}{2}$ for an $N$ color theory on a system with (lattice) volume $V$ {\it independent} of coupling $\lambda$.
For $SU(2)$ this implies $S_B=18.0$ for all $\lambda$.
The results are clearly
consistent with this prediction to a fraction of a percent as $\mu^2\to 0$ even for very large values of the coupling confirming the presence of an exact supersymmetry.
Even more important there is no sign of the phase transition that had been seen before in the $U(2)$ theory. Indeed all the observables we have looked at show smooth dependence on $\lambda$
providing evidence that the lattice theory possesses only a single phase
out to arbitrarily arbitrarily strong coupling.
It is interesting to note that the bosonic action is proportional to $N^2$ and not $N^2-1$ even though we suppress the $U(1)$ modes.  That is because they are still present in this formulation; rather than being removed, they are being tamed.  The new terms added to the action mostly affect the vacuum of these fields---which is why they still contribute to the counting of degrees of freedom.

\begin{figure}[htbp]
\centering
\includegraphics[width=0.75\textwidth]{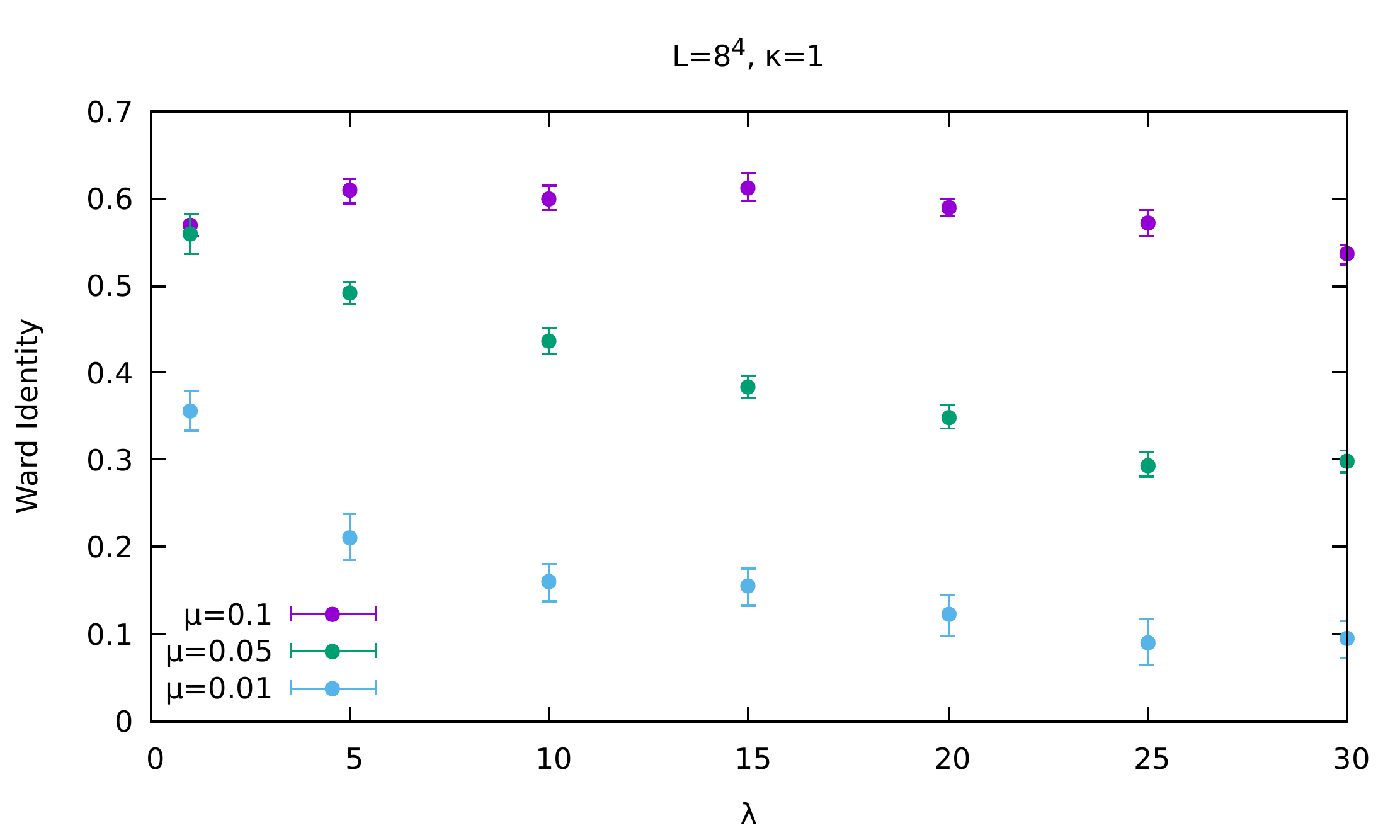} 
\caption{\label{bilinear}Bilinear $\cQ$-susy ward identity vs $\lambda$ for $8^4$ lattices for $\mu=0.1,0.05,0.01$}
\end{figure}

\begin{figure}[htbp]
\centering
\includegraphics[width=0.75\textwidth]{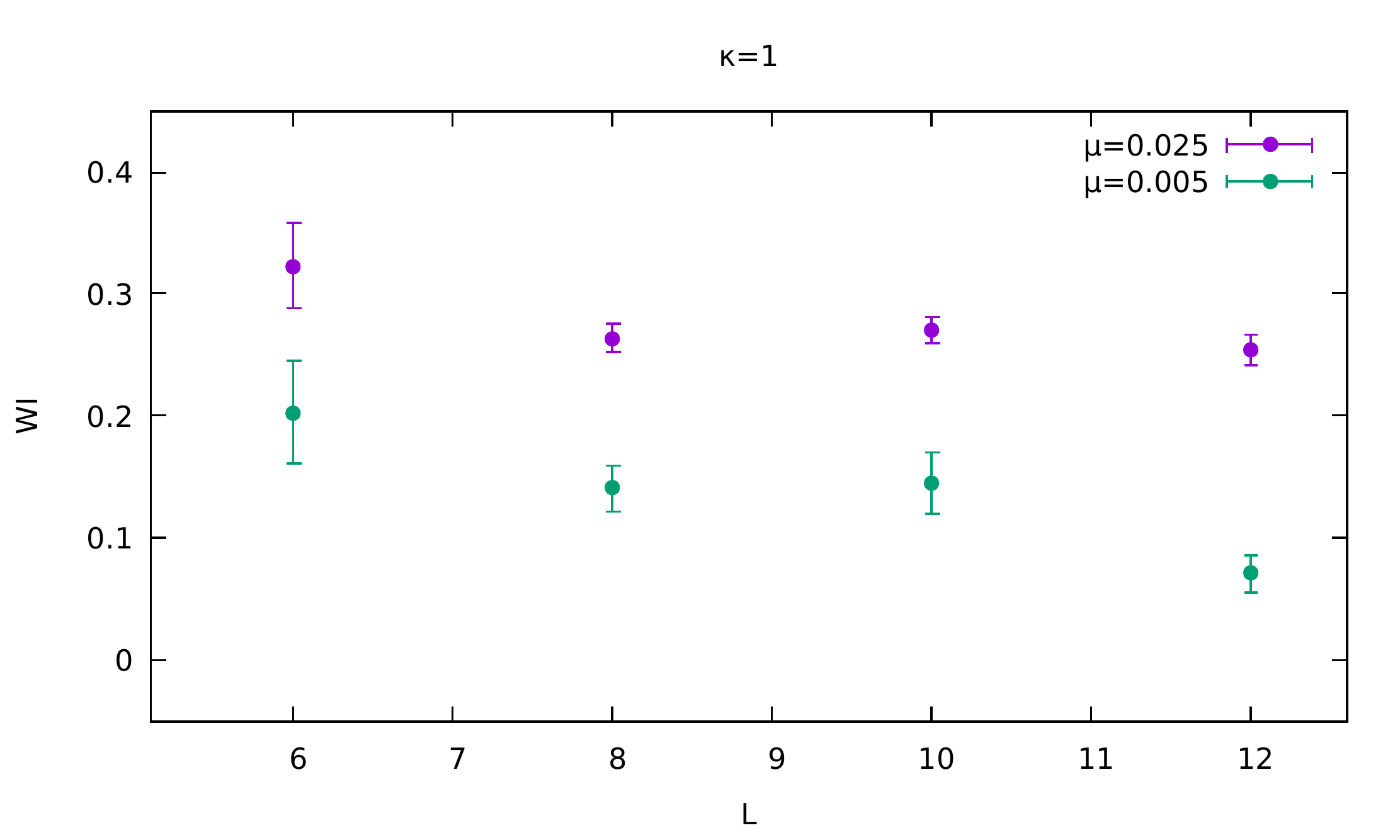} 
\caption{\label{ward_vs_L}Bilinear $\cQ$-susy ward identity vs $L$ at $\lambda=10.0$ for $\mu=0.025, 0.005$}
\end{figure}

Further confidence in this finding  comes
from studying a simple bilinear  Ward identity given by $\left\langle \cQ \Tr(\eta\cU_a\cUb_a)\right\rangle=0$. Fig.~\ref{bilinear} shows this quantity as a function
of $\lambda$ for several $\mu$ at $L=8$. It falls slowly with $\lambda$
and decreases more quickly with decreasing $\mu$. 
To clarify its dependence on lattice size we plot
the Ward identity for $\lambda=10.0$ vs $L$ for two values of $\mu$ in Fig.~\ref{ward_vs_L}. 
This plot makes it clear that the Ward identity decreases with increasing $L$. Indeed, comparing
$L=6$  with $L=12$ at $\mu=0.005$ the change is consistent
with a $1/L^2$ dependence on lattice size. 

\begin{figure}[htbp]
\centering
\includegraphics[width=0.75\textwidth]{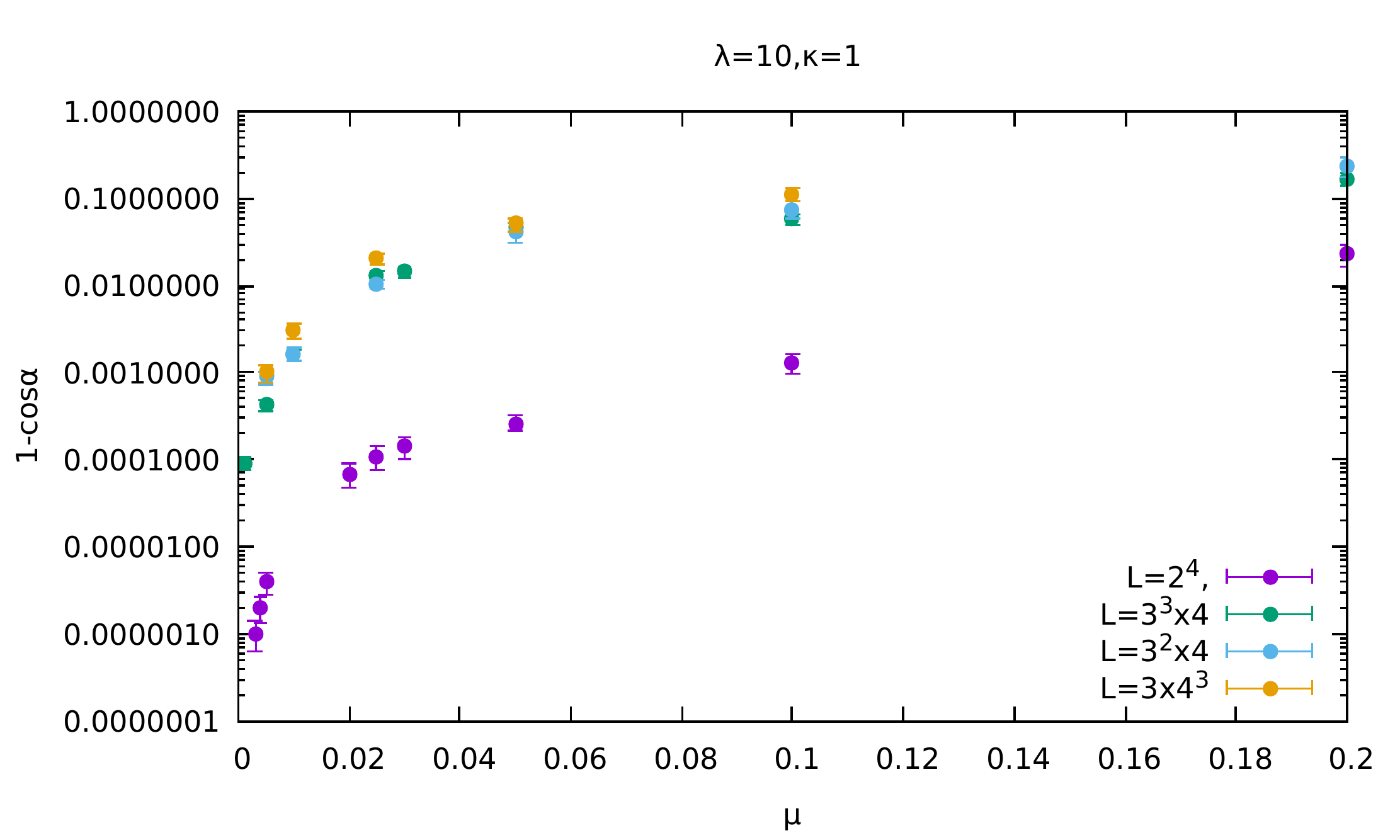} 
\caption{\label{pfaffl10} Pfaffian phase vs $\mu$ at $\lambda=10.0$ }
\end{figure}

\section{Absence of a sign problem}

Of course these results are derived from simulations of a model in which the phase of
the Pfaffian that results from fermion integration is neglected. To check for the presence
of such a phase we have computed it using the ensemble of configurations generated in our phase quenched Monte Carlo.
Writing the Pfaffian phase as $e^{i\alpha(\lambda,U)}$ we plot the quantity
$1-\cos{\alpha}$ as a function of $\mu$ at $\lambda=10.0$ and $\kappa=1.0$ in fig.~\ref{pfaffl10}. The different
data points 
correspond to lattices of size  
$2^4$, $3^2\times 4^2$, $3^3 \times 4$ and $3\times 4^3$ respectively. 
When measuring the phase of the Pfaffian we set $\kappa=0$ in the fermion
operator. 
Clearly the phase angle is driven towards very small values for small enough $\mu$. We have observed this for all values of $\lambda$  -- the analogous plot fig.~\ref{pfaffl30} for
$\lambda=30$ is shown in the appendix. Of course the lattices used in these tests are quite small
and one should worry whether the sign problem returns on larger volumes. Our results suggest that
this is not the case -- the average phase appears to saturate as the volume increases. Systems with sign problems typically exhibit phase fluctuations that increase exponentially with volume.
This lattice model seems very different in this regard.
\begin{figure}[htbp]
\centering
\includegraphics[width=0.75\textwidth]{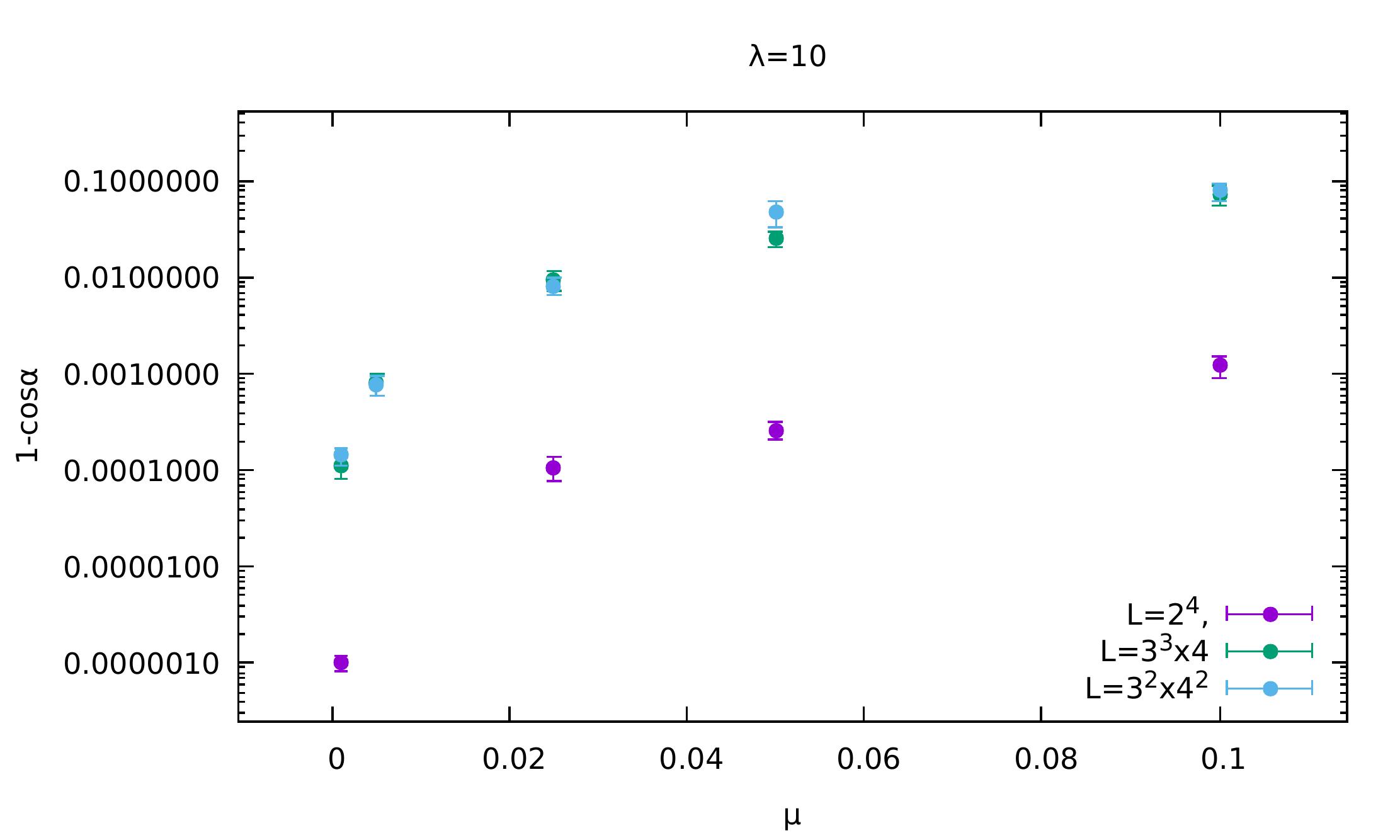} 
\caption{\label{newphase} Pfaffian phase vs $\mu$ at $\lambda=10.0$. Ensembles were generated without including the new fermionic term }
\end{figure}

Retaining the new $U(1)$ breaking fermion term in the evolution but neglecting it when measuring the
phase is clearly a questionable procedure.  However, the modification that is neglected relates to the trace modes, which decouple from the $SU(N)$ theory in the continuum in any case.  So, in some sense we are discarding an irrelevant piece.  Nevertheless, we have also generated ensembles in which the new fermion
term is dropped from the fermion action in both the evolution
and the measurement of the phase. A typical plot of the resultant
phase versus for $\mu$ at $\lambda=10.0$ is shown in fig.~\ref{newphase} 
for several lattice volumes. The observed
behavior is very similar to that seen in fig.~\ref{pfaffl10} and lends confidence to the assertion
that the system does not suffer from a sign problem. Since this procedure breaks $\cQ$-symmetry
softly (proportional to $\kappa$) it leads to larger deviations in the Ward identities and so we have reinstated the
new fermion term in our later simulations used for studying Wilson loops.  The fact that eliminating the new fermion term from both the Pfaffian measurement and the simulation still preserves the good behavior can be understood as the new bosonic term accomplishing the most important task:  stabilizing and suppressing the $U(1)$ modes of the link fields in a $\cQ$-symmetric way that is only softly broken.

It is interesting to try and understand theoretically why the observed phase fluctuations
are so small. We start by writing the expectation value of the phase measured in the phase
quenched ensemble as
\beq
<e^{i\alpha(\lambda,\kappa,U)}>_{\rm phase\, quenched}=\int D\cU D\cUb\, e^{i\alpha(\kappa,\lambda,\cU)} \vert{\rm Pf}(\cU)\vert e^{-S_B(\lambda,\kappa,\cU)}=1\eeq
where we have chosen the normalization of the measure so that the full partition with
susy preserving periodic boundary conditions (the Witten index) is unity.
Furthermore, $\cQ$-invariance ensures that
this expectation value of the phase factor is independent of $\kappa$ and can be 
computed for $\kappa\to\infty$ where the partition function is saturated by
configurations with unit determinant - the $SU(2)$ theory.
Finally, the topological character of this partition function can be exploited to localize
the integral to configurations which are constant over the lattice -- the integral reducing to a Yang-Mills matrix model integral. The resultant
Pfaffian for the $SU(2)$ matrix model is known to be real, positive definite  \cite{Krauth:1998xh}. 
Of course our simulations are performed
at finite $\kappa$, and use a thermal boundary condition, but the numerical results strongly suggest that as a practical matter the phase
fluctuations are small for the relevant range of parameters.

The encouraging results for the phase of the Pfaffian may also be related to the fact that out to very large $\lambda$ the center symmetry is unbroken, so that Eguchi-Kawai reduction \cite{Eguchi:1982nm} may be valid.  In that case the theory is equivalent to a single-site lattice, where the gauge theory is in fact just the matrix model that has been indicated in the previous paragraph.  This may also explain why we are able to obtain results consistent with large $N$ predictions (below), since the fact that we are in volumes larger than a single site may in fact translate into larger $N$ in the reduced model.

\section{Supersymmetric Wilson loops}

\begin{figure}[htbp]
\centering
\includegraphics[width=0.75\textwidth]{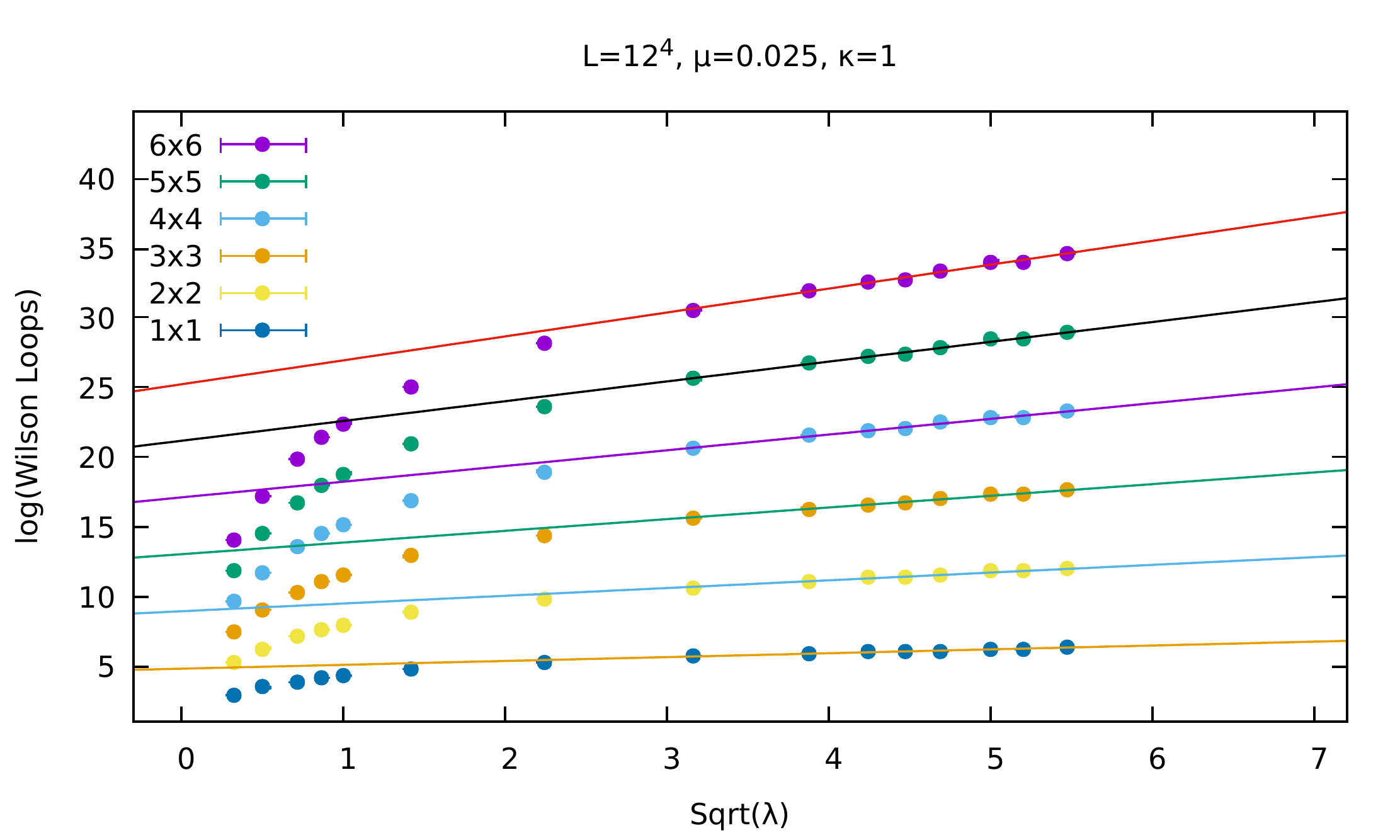} 
\caption{\label{loops}Supersymmetric $n\times n$ Wilson loops on $12^4$ lattice at $\mu=0.025$ }
\end{figure}

The previous results provide strong evidence that the lattice theory exists in a single phase with unbroken supersymmetry
out to very large values of the gauge coupling and that the model can be simulated with a Monte Carlo
algorithm without encountering a sign problem. With this in hand we turn to whether the lattice simulations
can provide confirmation of known results for $\cN=4$ Yang-Mills at strong coupling. Most of these analytic results were obtained
by exploiting the AdS/CFT correspondence which allows strong coupling results in the gauge theory to be obtained
by solving a classical gravity problem in anti-de Sitter space. Using this duality a variety of results for
{\it supersymmetric} Wilson loops  have been obtained over the last twenty years. Such Wilson loops generalize the usual Wilson loops by including contributions from
the scalars and are realized in the twisted construction by forming path ordered products of the {\it complexified} lattice gauge fields $\cU_a$. 
In the continuum  the generic feature of such Wilson loops is that for strong coupling they 
depend not on $\lambda$ as one would expect from perturbation theory but instead vary like $\sqrt{\lambda}$.
\begin{figure}[htbp]
\centering
\includegraphics[width=0.75\textwidth]{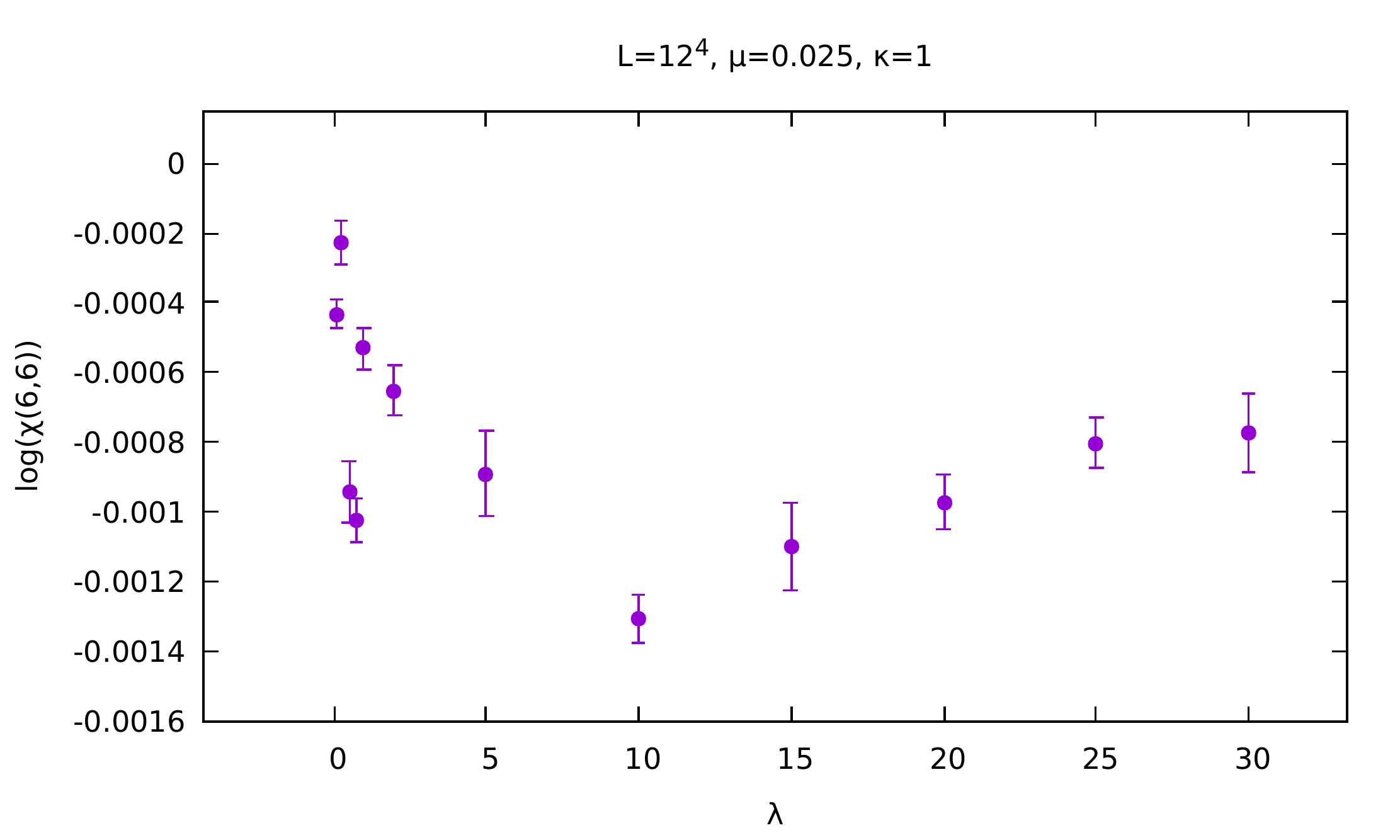} 
\caption{\label{creutz} Estimated string tension vs $\lambda$ for $12^4$ lattice at $\mu=0.025$ using
$\chi(6,6)$}
\end{figure}
In fig.~\ref{loops} we show the logarithm of the $n\times n$ supersymmetric Wilson loops $W(n,n)$
for a $12^4$ lattice at $\kappa=1.0$ plotted as a function of $\sqrt{\lambda}$. The straight lines correspond to fits with $\sqrt{\lambda}\ge 3$. It is clear that all the
loops show a $\sqrt{\lambda}$ dependence at strong coupling in agreement with the holographic prediction.
This is encouraging. It is also clear that the fits show a linear dependence on the length of the
perimeter of the loop.
If we parametrize the static potential defined by
$W(R,T)=e^{-V(R)T}$ in the form
\beq
V(R)=\sigma(\lambda) R+\alpha(\lambda)/R+M(\lambda)\eeq
The presence of the constant term 
$M(\lambda)$ will yield the observed perimeter scaling provided the string
tension is small or zero. Such a perimeter
term also occurs in continuum
treatments where it corresponds to the energy of a static probe 
source in the fundamental representation
and has to be explicitly subtracted out to see the non-abelian Coulomb behavior hidden
in $\alpha(\lambda)$ \cite{Erickson:2000af}. 

One way to remove the 
perimeter dependence is to consider Creutz ratios defined by
\beq
\chi(R,T)=\frac{W(R,T)W(R-1,T-1)}{W(R,T-1)W(R-1,T)}\eeq
For a theory with Wilson loops containing both perimeter, area and Coulomb behaviors
one finds 
\beq\ln{\chi(R,R)}\sim -\sigma(\lambda) + \alpha(\lambda)/R^2\eeq
Thus we can read of the string tension by examining the large $R$ behavior of $\ln{\chi(R,R)}$.
In fig.~\ref{creutz} we plot $\ln{\chi(6,6)}=-\sigma$ versus $\lambda$ for a $12^4$ lattice at $\lambda=10.0$ and $\mu=0.025$.

Clearly, the string tension is very small even at strong coupling
which is consistent with the existence of a single superconformal phase in the theory in the IR.
\begin{figure}[htbp]
\centering
\includegraphics[width=0.75\textwidth]{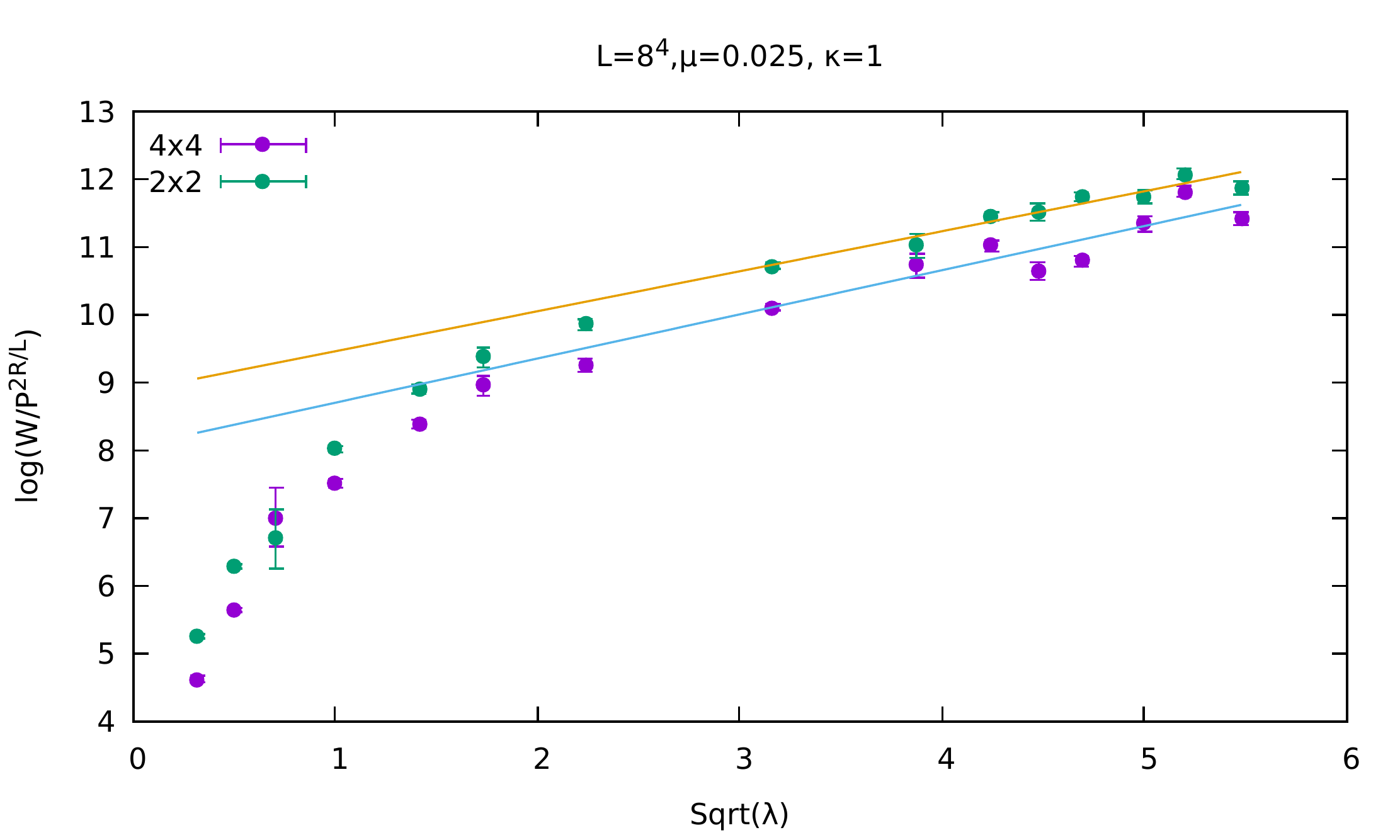} 
\caption{\label{wilson4} Renormalized supersymmetric $4\times 4$ and $2\times 2$ Wilson loops on $8^4$ lattice at $\mu=0.025$ }
\end{figure}
Of course the most interesting question is 
whether we can see evidence for a non-abelian Coulomb potential at small $R$. Direct fits to the Creutz ratio are consistent with the presence of such a term but the errors in $\alpha(\lambda)$ are large.

\begin{figure}[htbp]
\centering
\includegraphics[width=0.75\textwidth]{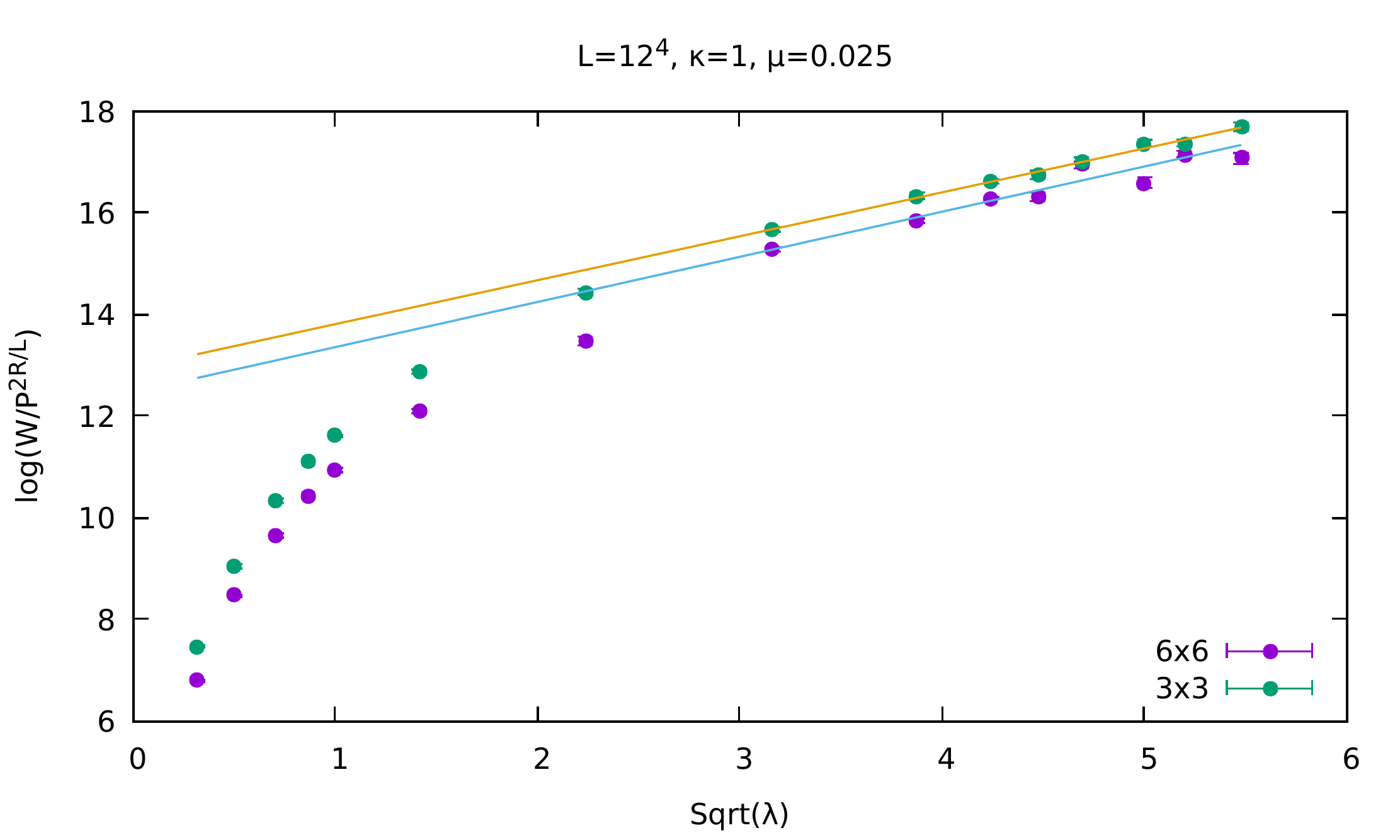} 
\caption{\label{wilson6}Renormalized supersymmetric $6\times 6$ and $3\times 3$ Wilson loops on $12^4$ lattice at $\mu=0.025$ }
\end{figure}

An alternative way to probe for this is
is to divide the original Wilson loops by an
appropriate power of the measured Polyakov line $P$ which is given by product of
gauge links along a thermal cycle. The (logarithm of the) Polyakov line also picks up a term linear
in the length of the lattice due to a massive source
and hence can used to subtract the linear divergence 
in the rectangular Wilson loop.
We thus define a renormalized Wilson loop on a
$L^4$ lattice of the form
\beq W^R(R,R)=\frac{W(R,R)}{P^{\frac{2R}{L}}}\eeq
These are shown
in fig.~\ref{wilson4} for a $8^4$ lattice.
Notice that the $2\times 2$ and $4\times 4$ loops now lie near to
each other which is consistent with conformal invariance and the presence of
a non-abelian Coulomb term
while the strong coupling behavior
still exhibits a dependence on $\sqrt{\lambda}$. 
This result can also be seen on the larger $12^4$ lattice shown in fig.~\ref{wilson6}.
Notice that the
average slope in this case is somewhat larger than the data on $8^4$. This presumably reflects 
the residual
breaking of conformal invariance due to finite volume as well as finite lattice spacing.
However it may also indicate that our definition of a 
renormalized Wilson loop does not do a perfect job of subtracting all the linear divergences
needed to reveal an underlying Coulombic term. Further work is needed on larger lattices to clarify this issue.

Details of the fits for the different Wilson loops and lattice are shown in tables \ref{wilsontable_l8_renorm},\ref{wilsontable_l12_renorm}.

The square root behavior at large
$\lambda$ is consistent with the result for circular Wilson loops in ${\cal N}=4$ SYM derived by Gross and Drukker \cite{Drukker:2000rr} and Maldacena's holographic argument \cite{Maldacena:1998im}. There are also explicit calculations using
holography for the rectangular Wilson loop in \cite{Erickson:2000af}.
The strange $\sqrt{\lambda}$ dependence {\it cannot} be seen in perturbation theory
and this (admittedly) very preliminary result is a very non-trivial test of the correctness of the lattice approach in a non-perturbative regime.
\begin{table}[htbp]
\centering
\begin{tabular}{|c|c|c|}
\hline
Loop Size & a$\sqrt{\lambda}$+b & Reduced-$\chi^2$ \\ \hline
$4\times4$              & 0.6(1)$\sqrt{\lambda}$ + 8.0(4)    & 8.11           \\ \hline
$2\times2$              & 0.59(4)$\sqrt{\lambda}$ + 8.8(2)    & 2.25          \\ \hline
\end{tabular}
\caption{\label{wilsontable_l8_renorm} Normalized Supersymmetric  Wilson loop fits  on $8^4$ lattice at $\mu=0.025$ for $f(\lambda)= a \sqrt{\lambda} + b$ }
\end{table}

\begin{table}[htbp]
\centering
\begin{tabular}{|c|c|c|}
\hline
Loop Size & a$\sqrt{\lambda}$+b & Reduced-$\chi^2$ \\ \hline
$6\times6$              & 0.88(7)$\sqrt{\lambda}$ + 12.4(3)    & 6.58           \\ \hline
$3\times3$              & 0.86(2)$\sqrt{\lambda}$ +12.94(9)    & 0.90           \\ \hline
\end{tabular}
\caption{\label{wilsontable_l12_renorm} Normalized Supersymmetric  Wilson loop fits  on $12^4$ lattice at $\mu=0.025$ for $f(\lambda)= a \sqrt{\lambda} + b$ }
\end{table}

\section{Conclusions}
We have found that a supersymmetric modification of the lattice action enables us to extend our simulations to what seem to be arbitrarily large values of the 't Hooft coupling without encountering difficulties that had previously limited our studies to modest $\lambda$.  This seems to be attributable to stabilizing the potential for the $U(1)$ modes in a way that preserves the essential ${\cal Q}$ supersymmetry of the construction.  The current study has been limited to gauge group $SU(2)$.  It is natural to inquire what occurs for this construction for other $SU(N)$.  We will investigate this in future studies; however, we expect that a sign problem will reemerge since in the zero-dimensional matrix models for $N>2$ the Pfaffian is no longer strictly positive.

\acknowledgments
This work was supported by the US Department of Energy (DOE), Office of Science, Office of High Energy Physics, 
under Award Numbers {DE-SC0009998} (SC,GT) and {DE-SC0013496} (JG). Numerical calculations were carried out on the DOE-funded USQCD facilities at Fermilab.
The authors would like to thank David Schaich for help with the parallel code  used in this work. 
\newpage
\appendix
\section{Appendix}

\begin{figure}[htbp]
\centering
\includegraphics[width=0.75\textwidth]{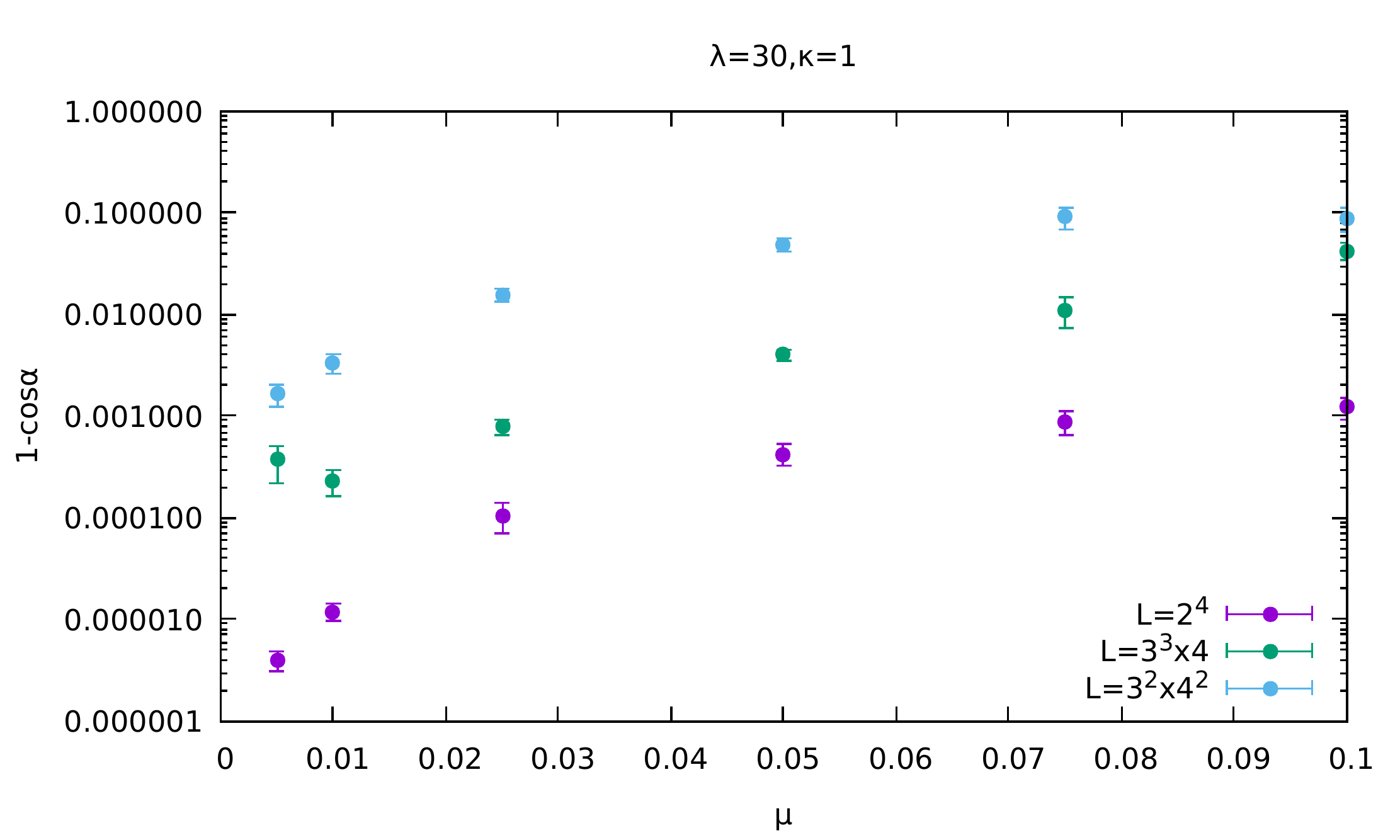} 
\caption{\label{pfaffl30} Pfaffian phase vs $\mu$ at $\lambda=30.0$ }
\end{figure}

\bibliographystyle{JHEP3}
\bibliography{susylink}

\providecommand{\href}[2]{#2}\begingroup\raggedright\begin{thebibliography}{10}

\bibitem{Catterall:2009it}
S.~Catterall, D.~B. Kaplan and M.~Unsal, \emph{{Exact lattice supersymmetry}},
  \href{https://doi.org/10.1016/j.physrep.2009.09.001}{\emph{Phys. Rept.}
  {\bfseries 484} (2009) 71} [\href{https://arxiv.org/abs/0903.4881}{{\ttfamily
  0903.4881}}].

\bibitem{Catterall:2011pd}
S.~Catterall, E.~Dzienkowski, J.~Giedt, A.~Joseph and R.~Wells,
  \emph{{Perturbative renormalization of lattice N=4 super Yang-Mills theory}},
  \href{https://doi.org/10.1007/JHEP04(2011)074}{\emph{JHEP} {\bfseries 1104}
  (2011) 074} [\href{https://arxiv.org/abs/1102.1725}{{\ttfamily 1102.1725}}].

\bibitem{Catterall:2012yq}
S.~Catterall, P.~H. Damgaard, T.~Degrand, R.~Galvez and D.~Mehta, \emph{{Phase
  Structure of Lattice N=4 Super Yang-Mills}},
  \href{https://doi.org/10.1007/JHEP11(2012)072}{\emph{JHEP} {\bfseries 1211}
  (2012) 072} [\href{https://arxiv.org/abs/1209.5285}{{\ttfamily 1209.5285}}].

\bibitem{Catterall:2013roa}
S.~Catterall, J.~Giedt and A.~Joseph, \emph{{Twisted supersymmetries in lattice
  ${\cal N}=4$ super Yang-Mills theory}},
  \href{https://doi.org/10.1007/JHEP10(2013)166}{\emph{JHEP} {\bfseries 1310}
  (2013) 166} [\href{https://arxiv.org/abs/1306.3891}{{\ttfamily 1306.3891}}].

\bibitem{Catterall:2014vka}
S.~Catterall, D.~Schaich, P.~H. Damgaard, T.~DeGrand and J.~Giedt, \emph{{N=4
  Supersymmetry on a Space-Time Lattice}},
  \href{https://doi.org/10.1103/PhysRevD.90.065013}{\emph{Phys. Rev. D}
  {\bfseries 90} (2014) 065013}
  [\href{https://arxiv.org/abs/1405.0644}{{\ttfamily 1405.0644}}].

\bibitem{Schaich:2018mmv}
D.~Schaich, \emph{{Progress and prospects of lattice supersymmetry}},
  \href{https://doi.org/10.22323/1.334.0005}{\emph{PoS} {\bfseries LATTICE2018}
  (2019) 005} [\href{https://arxiv.org/abs/1810.09282}{{\ttfamily
  1810.09282}}].

\bibitem{Anagnostopoulos:2007fw}
K.~N. Anagnostopoulos, M.~Hanada, J.~Nishimura and S.~Takeuchi, \emph{{Monte
  Carlo studies of supersymmetric matrix quantum mechanics with sixteen
  supercharges at finite temperature}},
  \href{https://doi.org/10.1103/PhysRevLett.100.021601}{\emph{Phys. Rev. Lett.}
  {\bfseries 100} (2008) 021601}
  [\href{https://arxiv.org/abs/0707.4454}{{\ttfamily 0707.4454}}].

\bibitem{Hanada:2008gy}
M.~Hanada, A.~Miwa, J.~Nishimura and S.~Takeuchi, \emph{{Schwarzschild radius
  from Monte Carlo calculation of the Wilson loop in supersymmetric matrix
  quantum mechanics}},
  \href{https://doi.org/10.1103/PhysRevLett.102.181602}{\emph{Phys. Rev. Lett.}
  {\bfseries 102} (2009) 181602}
  [\href{https://arxiv.org/abs/0811.2081}{{\ttfamily 0811.2081}}].

\bibitem{Catterall:2008yz}
S.~Catterall and T.~Wiseman, \emph{{Black hole thermodynamics from simulations
  of lattice Yang-Mills theory}},
  \href{https://doi.org/10.1103/PhysRevD.78.041502}{\emph{Phys. Rev. D}
  {\bfseries 78} (2008) 041502}
  [\href{https://arxiv.org/abs/0803.4273}{{\ttfamily 0803.4273}}].

\bibitem{Catterall:2009xn}
S.~Catterall and T.~Wiseman, \emph{{Extracting black hole physics from the
  lattice}}, \href{https://doi.org/10.1007/JHEP04(2010)077}{\emph{JHEP}
  {\bfseries 04} (2010) 077} [\href{https://arxiv.org/abs/0909.4947}{{\ttfamily
  0909.4947}}].

\bibitem{Catterall:2010fx}
S.~Catterall, A.~Joseph and T.~Wiseman, \emph{{Thermal phases of D1-branes on a
  circle from lattice super Yang-Mills}},
  \href{https://doi.org/10.1007/JHEP12(2010)022}{\emph{JHEP} {\bfseries 12}
  (2010) 022} [\href{https://arxiv.org/abs/1008.4964}{{\ttfamily 1008.4964}}].

\bibitem{Hanada:2016zxj}
M.~Hanada, Y.~Hyakutake, G.~Ishiki and J.~Nishimura, \emph{{Numerical tests of
  the gauge/gravity duality conjecture for D0-branes at finite temperature and
  finite N}}, \href{https://doi.org/10.1103/PhysRevD.94.086010}{\emph{Phys.
  Rev. D} {\bfseries 94} (2016) 086010}
  [\href{https://arxiv.org/abs/1603.00538}{{\ttfamily 1603.00538}}].

\bibitem{Berkowitz:2016jlq}
E.~Berkowitz, E.~Rinaldi, M.~Hanada, G.~Ishiki, S.~Shimasaki and P.~Vranas,
  \emph{{Precision lattice test of the gauge/gravity duality at large-$N$}},
  \href{https://doi.org/10.1103/PhysRevD.94.094501}{\emph{Phys. Rev. D}
  {\bfseries 94} (2016) 094501}
  [\href{https://arxiv.org/abs/1606.04951}{{\ttfamily 1606.04951}}].

\bibitem{Catterall:2017lub}
S.~Catterall, R.~G. Jha, D.~Schaich and T.~Wiseman, \emph{{Testing holography
  using lattice super-Yang-Mills theory on a 2-torus}},
  \href{https://doi.org/10.1103/PhysRevD.97.086020}{\emph{Phys. Rev. D}
  {\bfseries 97} (2018) 086020}
  [\href{https://arxiv.org/abs/1709.07025}{{\ttfamily 1709.07025}}].

\bibitem{Rinaldi:2017mjl}
E.~Rinaldi, E.~Berkowitz, M.~Hanada, J.~Maltz and P.~Vranas, \emph{{Toward
  Holographic Reconstruction of Bulk Geometry from Lattice Simulations}},
  \href{https://doi.org/10.1007/JHEP02(2018)042}{\emph{JHEP} {\bfseries 02}
  (2018) 042} [\href{https://arxiv.org/abs/1709.01932}{{\ttfamily
  1709.01932}}].

\bibitem{Catterall:2014vga}
S.~Catterall, J.~Giedt, D.~Schaich, P.~H. Damgaard and T.~DeGrand,
  \emph{{Results from lattice simulations of N=4 supersymmetric Yang--Mills}},
  {\emph{PoS} {\bfseries LATTICE2014} (2014) 267}
  [\href{https://arxiv.org/abs/1411.0166}{{\ttfamily 1411.0166}}].

\bibitem{Catterall:2015ira}
S.~Catterall and D.~Schaich, \emph{{Lifting flat directions in lattice
  supersymmetry}}, \href{https://doi.org/10.1007/JHEP07(2015)057}{\emph{JHEP}
  {\bfseries 07} (2015) 057}
  [\href{https://arxiv.org/abs/1505.03135}{{\ttfamily 1505.03135}}].

\bibitem{Krauth:1998xh}
W.~Krauth, H.~Nicolai and M.~Staudacher, \emph{{Monte Carlo approach to M
  theory}}, \href{https://doi.org/10.1016/S0370-2693(98)00557-7}{\emph{Phys.
  Lett. B} {\bfseries 431} (1998) 31}
  [\href{https://arxiv.org/abs/hep-th/9803117}{{\ttfamily hep-th/9803117}}].

\bibitem{Eguchi:1982nm}
T.~Eguchi and H.~Kawai, \emph{{Reduction of Dynamical Degrees of Freedom in the
  Large N Gauge Theory}},
  \href{https://doi.org/10.1103/PhysRevLett.48.1063}{\emph{Phys.Rev.Lett.}
  {\bfseries 48} (1982) 1063}.

\bibitem{Erickson:2000af}
J.~Erickson, G.~Semenoff and K.~Zarembo, \emph{{Wilson loops in N=4
  supersymmetric Yang-Mills theory}},
  \href{https://doi.org/10.1016/S0550-3213(00)00300-X}{\emph{Nucl. Phys. B}
  {\bfseries 582} (2000) 155}
  [\href{https://arxiv.org/abs/hep-th/0003055}{{\ttfamily hep-th/0003055}}].

\bibitem{Drukker:2000rr}
N.~Drukker and D.~J. Gross, \emph{{An Exact prediction of N=4 SUSYM theory for
  string theory}}, \href{https://doi.org/10.1063/1.1372177}{\emph{J. Math.
  Phys.} {\bfseries 42} (2001) 2896}
  [\href{https://arxiv.org/abs/hep-th/0010274}{{\ttfamily hep-th/0010274}}].

\bibitem{Maldacena:1998im}
J.~M. Maldacena, \emph{{Wilson loops in large N field theories}},
  \href{https://doi.org/10.1103/PhysRevLett.80.4859}{\emph{Phys. Rev. Lett.}
  {\bfseries 80} (1998) 4859}
  [\href{https://arxiv.org/abs/hep-th/9803002}{{\ttfamily hep-th/9803002}}].

\end{thebibliography}\endgroup

\end{document}